# How does 'undone science' get funded? A bibliometric analysis linking rare diseases publications to national and European funding sources

Alex Rushforth \*, Alfredo Yegros-Yegros\*\*, Philippe Mongeon \*\*, Thed van Leeuwen\*\*

Corresponding author: alexander.rushforth@phc.ox.ac.uk; leeuwen@cwts.nl; a.yegros@cwts.leidenuniv.nl

\* Nuffield Department of Primary Care Health Sciences,
University of Oxford
Radcliffe Observatory Quarter, Woodstock Road
Oxford. OX2 6GG, UK

\*\* Center for Science & Technology Studies (CWTS)
Leiden University
Kolffpad 1
Leiden, the Netherlands

**Cite this as**: Rushforth, A. D., Yegros, A., Mongeon, P. & van Leeuwen, T. (2016). How does 'undone science' get funded? A bibliometric analysis linking rare diseases publications to national and European funding sources. EU-SPRI Early Career Researcher Conferences. Vienna, November 21, 2016

## Introduction

Strains in the linear model of innovation since in the 1980s and 1990s and dwindling block funding support for research have created increasing pressures on public sciences to respond to needs and demands articulated by public policy and other societal actors. The increased propensity for stakeholders of public science to 'speak back' was of course noted in discussion around 'Mode 2' knowledge production (Nowotny et al., 2001). Similarly 'triple helix' and 'academic capitalism' literatures observed the growing influence of private actors in funding and collaborating with university researchers and considered the effects this would have in shaping research agendas and academic norms (Etzkowitz and Leydesdorff, 2000, Slaughter and Leslie, 1997). Recent years have seen an increasing embedding of efforts to ensure science delivers on its promises of social relevance, with demands being inscribed into instruments of science policy like funding, evaluation and priority setting. In light of calls for science to become more 'context sensitive', one of the most urgent questions for science policy is who's definitions of societal needs and 'relevance' come to be heard (Sarewitz and Pielke, 2007, Rushforth and de Rijcke, 2017). Whilst interests articulated by government and by industry are certainly likely to persist, recent accounts in science studies (STS) have noted the increasing influence of civil society and social movement groups in steering the course of public policy and even directly influencing and funding scientific research. This is particularly visible around issues of undone science whereby "areas of research [are] identified by social movements and other civil society organizations as having potentially broad social benefit that are left unfunded, incomplete, or generally ignored" (Frickel et al., 2010, 445).

One of the notable features of undone science debates is how formation of new interest groups becomes pivotal in mobilizing and championing emerging research on undone topics. Clearly money is one of the most important mediums through which different types of actors can support and steer scientists to work on undone topics. Yet which actors are more visible in their support for scientific research is something which has seldom been measured. This study delves into research funding in the context of rare diseases research – a topic which has evolved from the margins of medical research into a priority area articulated by many contemporary funding agencies. Rare diseases refer to conditions affecting relatively few people in a population. Given low incidences, interest groups have articulated a lack of attention within medical research compared to more common conditions. The rise to prominence of rare diseases in research funding policies is often explained in the science studies literature in terms of effective lobbying by social movements (especially patient advocate organisations), persuading policymakers to effect changes in research priorities (Callon and Rabeharisoa, 2003, Rabeharisoa, 2003, Rabeharisoa, 2006). Likewise, innovative fundraising initiatives, infrastructure building, and close partnerships with research groups are other means through which interested actors have sought to build capacity for research into rare medical conditions. To date however systematic empirical evidence to compare the relative importance of different actors in funding rare disease research has not been produced. Building on interest in undone science in STS and science policy studies, our study hopes to map-out different kinds of funding actors and their influence on leading scientific research on rare diseases, by use of bibliometric tools. The approach we are developing relies on the use of Funding Acknowledgement (FA) data provided in Thomson Reuters Web of Science database.

We analyze different kinds of funding sources linked the 'high impact' and 'scientifically leading' research in rare disease research over a fixed period (2009-2015) within four EU member states: UK, Netherlands, France, and Spain. Here we take publication output and citation impact analysis as a logical – if not exhaustive – criteria for evaluating the presence and importance of different kinds of funding actors in support of this topic. Using these outcome indicators, we draw some tentative conclusions into the means through which building capacity for rare disease research seems to have been built in these four countries. These are relevant cases because their national scientific research councils have each committed to building scientific capacity by funding scientific projects and consortia, selected on the basis of quality via peer review.

The current analysis presented in this manuscript is subject to certain limitations at the present time. Both the dataset itself and the analysis we perform in this paper should be treated as 'work in progress', as further data cleaning and steps in our analysis need to be taken before our results can be presented as fully reliable. For the purpose of this paper however, we lay out the objectives of our study, the steps we have taken so far, and the next steps we hope to carry out to realise the research objectives. In the following section we provide an outline of the rise to prominence of rare diseases as a priority topic in European science policy, before discussing the strengths and limitations of the approach we are taking. Next we lay out the methodological steps we have currently taken, followed by the (provisional) analysis and a fuller elaboration of the next steps to be taken.

## The rise of rare diseases in European research policy

Despite evident priority efforts, there is no single medical definition of rare diseases agreed-upon globally, as the term is region specific (which creates several methodological challenges for identifying rare disease research, see Methods section below). In the European Union a condition is defined as rare when it affects less than 1 in 2000 of the EU population, with an estimated 7000 such diseases discovered to date. Although single instances of rare diseases are uncommon, together they will directly affect around 6-8% of the European population during their lifetimes (approximately 27-36 million people) (EUROPLAN, 2009)[1]. Indeed some have argued that rare diseases should be considered a 'boundary object', that is a term of which the meaning is plastic and flexible enough to meet the interests, requirements, and understandings of several different groups (Huyard, 2009, 464). This definition is helpful as it helps to understand how a variety of actors have come to converge around an otherwise disparate set of medical conditions, and how these conditions gathered sufficient strength in numbers to become a priority area within European funding programs. Gulbrandsen et al (2016) recently remarked that terms like 'rare diseases' (as well as meta-priorities like 'grand challenges' or 'national security') represent politically acceptable slogans through which public policymakers seek to steer all parts of research and innovation processes (Gulbrandsen et al., 2016, 1494). Part of the common legitimation they note for funding rare diseases and orphan drugs rests upon assumptions of 'system' or 'market failure', which would otherwise remain marginal to priorities set within scientific communities and wider innovation systems (c.f. Mazzucato, 2015). When pooled together, the policy promises carried under this label include: a) to tackle a major (largely unmet) public health issue in the context of national healthcare systems (Ayme and Rodwell, 2011), and b) to stimulate economic growth through pharmaceutical development of 'orphan drug' products (Moors and Faber, 2007).

---

[1] Another report estimated that worldwide 400 million people could suffer during their lifetimes (WHO 2004).

As a collective social movement, 'rare disease' advocates have been remarkably effective at persuading European policymakers of the importance of this cause. Since 1994, rare diseases have featured in open funding calls of successive European Framework packages (beginning with FP4), for instance being listed as a priority theme within Challenge 1 of the FP7 program (Health: Personalized Medicine). Although not covered by our analysis, it is worth noting that the theme of rare diseases is still current in the Horizon 2020 program (notably the Health, Demographic Change and Wellbeing priority theme). A specific ERA-NET program oriented to synthesizing national and regional rare disease projects was also set-up in the first round of the Horizon 2020, under the title of E-RARE (although this did not continue into the second Horizon 2020 round).

National funding agencies efforts to fund rare diseases are also widespread across European Members States, which have been influenced to greater or lesser degrees by 'bottom-up' initiatives to respond to public health policy demands and by more 'top-down' coordinating efforts of EU policymaking. Templates for 'National Plans' (or 'National Strategies') for Rare Diseases were outlined by the European Council in 2009, with this model being subsequently rolled-out in member states (although not all member states have yet introduced plans or strategies). Whereas some 'early adopter' countries had plans in place prior to 2009 (France, Greece, Portugal), other states' national research councils have subsequently developed following the rollout of national plans. Other member states have still not made rare diseases an explicit priority focus. Due to visible efforts to prioritize funding of rare diseases at a national level, this study will focus on four member states which have established priority funding and national plans for rare diseases: France, Netherlands, Spain and the United Kingdom.

**Citation Analysis and Funding Acknowledgements**

The importance of funding for the production of scientific work has long been observed in science studies (Knorr-Cetina, 1982). Given changes in the funding and organization of public sciences, researchers in public sector organisations like universities have become increasingly dependent on external funds in many contexts. In many research fields and university systems external funding has become a necessary condition to do research at all (Laudel, 2006). Thus, the role of funding agencies in shaping knowledge production in the science system can be expected to have increased, especially in countries where funding agencies and universities have traditionally been strong (Whitley 2010).

Whilst certain bibliometric measures like the journal impact factor are (controversially) being enrolled in the evaluation and management of academic performance (Rushforth and de Rijcke, 2015), some quantitative measures may nonetheless provide useful means of gaging analytically the influence of research funding on the cognitive development of science. For example, reviews of publications have long served as a means of evaluating new knowledge created by recipients of grants (Cozzens and Melkers, 1997, Wagner and Alexander, 2013, van Leeuwen and Moed, 2012), how grants effect knowledge output and impact of scientists before and after they receive funds (Campbell et al., 2010), and how receipt of grants correlate with accumulation of citations over the course of scientific careers (Bloch et al., 2014, Mongeon et al., 2016).

This pursuit is however prone to some methodological limitations. Bibliometric evaluations of funding agencies and channels have often needed to triangulate bibliometric outputs of known grantees with interviews and survey data- a process which can be expensive, time consuming, and still prone to incomplete information. This has effectively meant studies have been restricted to focusing on the citation impact of individual research 'oeuvres' of known grantees, for instance comparing them with control groups like non-grantee researchers. This produces only broad-picture correlations, as direct links between publications and specific grants or funding agencies which may have supported those outcomes has been lacking. At best this means making only very general conclusions, such as scientists with grants tend to have higher citation impact than those without.

In this paper, we overcome some of these limitations by using funding acknowledgement (FA) data available in Thomson Reuters' Web of Science (WoS). One of the main promises of funding acknowledgement data is in being able to deliver a more standardized and reliable way of linking publications to funding schemes and funders. However, there are at present some limitations associated with the use of FA data provided by this database. Most importantly, Thomson Reuters only started to index FA data in the WoS in mid-2008, and only for articles included in the Science Citation Index. FA for the Social Sciences Citation Index (SSCI) are available for the most recent publications, since 2015. However, our analysis will cover publications from 2009 onwards and mainly publications included in the SCI. Paul-Hus et al. (2016) found important difference in the coverage of FA in the WoS based on the language of the publication. They found that while 44% of English language articles included funding FA between 2009 and 2015, it was the case for 35.9% and 7.6% of papers in Chinese and Korean, respectively, and for less than 1% of papers published in other languages. Since our study includes research from Spain, the Netherlands and France, we may be missing funding information for articles published in those countries' national languages. However, this limitation is also not likely to have an important effect on the results of our study since even in these countries, publishing in English is the norm, especially in the biomedical field (Van Leeuwen et al., 2001).. Finally details on how Thomson Reuters have constructed the algorithm for sorting their FA data are not yet transparent, meaning the 'completeness' of this data is difficult to assess.

Another more fundamental limitation of the use of FA to link funding agencies and published articles is that there may be cases where researchers neglected to include FA in the article (Costas and Leeuwen, 2012). Also sometimes researchers may decide to include a FA even if the publication is not closely related to the funding received with the intention of showing that the funding led to scientific results. Overall, however, the use of FA data from WoS reduces false positives by allowing a more precise and direct linkage of funding agencies, specific grants, and published articles.

In this paper we use bibliometric methods to assess the influence different types of funding sources on rare disease research in terms of output and citation impact. We hypothesize that an increased focus on rare disease by funding policies will lead to increased research on the topic, which can be measured using publication counts. We also use citations as a proxy for the scientific impact or influence of individual publications. Despite being standard practice in bibliometrics, such a use of citations still has some limitations. Most importantly, there are numerous factors other than a given work's quality and/or relevance that determine the number of citations, such as the topic of the research, the time of publication, and the reputation of the authors (Bornmann and Daniel, 2008). For this reason, normalized citation counts are usually preferred to raw counts when attempting to compare the scientific impact of different publications or group of publications. The normalized citation count of a paper is calculated by dividing the number of citations of that article by the average number of citations published in the same year in the same field.

As well as methodological problems we have identified, bibliometric studies of cognitive impact of funding on scientific outputs face certain conceptual hurdles. Although external funding is no doubt the 'oxygen' of research (particularly in fields of biomedical and health sciences with high start-up and instrumentation costs), there have always been other layers of epistemic governance which are known to shape research outcomes, including employment organisations (e.g. universities) and epistemic communities in which researchers are embedded (c.f. Whitley 2010). Determining direct causal links between funding sources and publication and citation outcomes is thus a difficult, if not impossible task. Nonetheless, whilst we recognize these intervening variables are always likely to persist, recent turns towards developing funding acknowledgement data infrastructures promises to help specify with greater precision how specific research outcomes (e.g. indexed publications) correlate with specific funding organisations, funding channels, and individual grants. Although at present there are a number of technical limitations and conceptual debates around the exact usefulness of FA data, we argue linking this data to publications has the advantage of drilling down into the efficacy of different research funding organisations' models and initiatives for facilitating influential work. As such, with the advent of FA data in bibliometrics we have reasons to be hopeful that more precise and accurate correlations may well be forthcoming between research outcomes and funding actors.

## Method

To collect rare disease papers, we use a list of known rare disease names from the European Commission's Orphanet registry. This database was established in partnership with the French funding agency INSERM in order to provide a resource for patients, physicians and researchers about different rare disorders.

Orphanet has become a key node in European rare disease networks, with the establishment of the database, but also a dedicated journal, the Orphanet Journal of Rare Diseases. The database is updated annually by clinical experts and information scientists, and represents a concerted effort to collate and classify different kinds of information relevant to rare diseases in Europe.

One obvious limitation to our study is that using the Orphanet dataset means that we only take the European Union's legal definition of rare diseases (1 in 2000 of the European population) into account when searching. Whereas this classification is based on prevalence of diseases in the European Union population, the United States for instance classifies rare diseases differently, according to incidences within its national population (fewer than 200,000). Whilst one can expect quite considerable overlap between European and US definitions, the exact overlap is not known and there is likely to be some variation across the regions. There are also different definitions in countries like Japan. Given rare diseases are by definition legalistic and region specific, it is not possible to generate a definitive global list of rare diseases. Due to its region-specificity, the list also presents some diseases as rare which may initially appear counter-intuitive. If we take the example of Malaria, this is classified as rare within the Orphanet list, yet clearly in other parts of the globe it is much more prevalent.

Despite these limitations, Orphanet provides an extensive and well-maintained resource for exploring what are considered rare diseases in one of the regions of the world clearly most equipped and motivated to build research capacity in this area. It has been claimed in several publications that the list is becoming a de facto standard for classifying rare diseases in many medical and research communities (BERNAUER, 2014, de Vrueh, 2014).

One of the database's features most relevant for our efforts to retrieve 'rare disease' research literature is the list of names of disorders supplied by the dataset. Our dataset was made by searching each of the rare disease disorders in titles, abstracts and keywords of articles and reviews published by researchers in France, Netherlands, Spain and UK between 2009 and 2015 and included the Clarivate Analytics' Web of Science. In order to ensure that these publications are related to biomedical research, the search was limited to 84 out of the 250 subject categories related to biomedical fields

From the original Orphanet list of rare disorders we did not consider synonyms for our search. Whilst this means our current data set is incomplete, this precaution was necessary in order to avoid including false positives in the dataset as an important number of synonyms retrieve irrelevant results, for example 'SPERM' or 'GET'. In the next steps of research we plan to include a disambiguated list of synonyms checked by Orphanet NL.

**Constructing a Typology of Funding Organisations**

Using the funding acknowledgements included by the authors in their publications, We constructed a typology of funding actors partly based on ongoing classification work performed on the CWTS in-house database (van Honk et al., 2016). At present this dataset has not been fully cleaned, yet in the coming months once cleaned we plan to make a more fine-grained classification of funding source types based on purposeful selection of search terms. These subsequent plans are referred to in this section alongside what we have done so far, and are returned to in more depth in the final section of this paper.

- We classified the publications on rare diseases produced by France, Netherlands, Spain and UK. These publications were then sorted into five categories: 'European-funded', 'National-funded', 'National and European funded', 'Other funding' and 'non-funded.

In this preliminary analysis, for 'national' funding we selected names of organisations from our FA dataset which have been classified within the CWTS in-house database. Included in this category are mainly public funding organizations with heterogeneous instruments set-up to encompass different goals, with different selection models, project structures (individual or team-based grants), time durations, and amounts of money. This is clearly not ideal, and in subsequent rounds of analysis we aim to refine this 'national' category, in order to distinguish between national funding agencies (e.g. MRC, INSERN, NWO) and charitable foundations (e.g. Wellcome Trust, Cancer Research UK, Dutch Heart Foundation, AFM Telefon).

The European-funded category is at present also heterogeneous, as it includes European Commission Framework Programs, European regional development funds, and also pan-European funding agencies like EMBO. In future analyses we plan to drill-down into the quantities of publications linked with European Framework Packages and measure their relative citation impact in relation to abovementioned national funding sources.

Given limits in the FA-dataset, our analysis will only cover publications that can be linked to the European Commission's Framework Programs covered under the period 2009-2015. This means our publication dataset will likely cover the bulk of the FP7 period, and the tail-end of FP6, however publications from our Horizon 2020 dataset have not yet been classified.

Clearly receipt of funding from European and national funding organisations is not mutually exclusive, as some (though relatively few) publications acknowledge funding in both categories. A third category has thus been included in our analysis to capture these overlapping publications ('National and European funded').

A fourth category has been labeled as 'Other funding' and it refers to those publications including an acknowledgement from a foreign public funding body or from a funding body for which the country is unknown, like companies[2] (e.g. GSK, Pfizer).'.

Finally a (tentative) benchmark against which to measure the outputs and impact of European and national funding are publications which do not include funding acknowledgments. This is of course a somewhat ambiguous category, as it is not clear whether the papers in fact received no external funding or whether the authors simply neglected to include the information.

Having classified our dataset according to these five provisional categories, we performed a bibliometric analysis on all publications in the dataset, in order to compare publication outputs and citation impact across papers linked to different funding sources.

In our analysis we will cover the following:

1) Numbers of publications per year by type of funding in the four EU Member State (MS)
2) Numbers of publications per year overall, per EU MS, by the type of funding
3) Impact analysis of the output overall per MS, and impact analysis of each MS in the three types of funding

**Results**

In this section we present the outcomes of the first analyses on the dataset on rare diseases research. For the four selected countries we will show the overall output and citation impact scores, for all the publications as well as for the various funding sources (as indicated above). We present two indicators in the following, the number of publications (denoted by P), as well as the field normalized impact, captured using mean normalized citation score ((MNCS) (Waltman et al, 2012). With respect to the latter indicator it is important to realize that it is a ratio score, comparing actual versus expected impact, which means that the value is meaningful, denoting worldwide average impact level.

In Figure 1, the overall annual output scores are displayed for the years 2009-2015. Among the four countries, we observe that UK covers the largest share of research on rare diseases, followed by France, with the Netherlands and Spain having largely similar output numbers.

---

[2] In the funding acknowledgements there is no information about the city or country where the funding organization is located. For many public funding organizations it is possible to identify to which country belongs, however in most of the cases is not possible to determine the location of companies, especially if they are multinational companies.

These output numbers more or less reflect the overall size differences between these four countries.

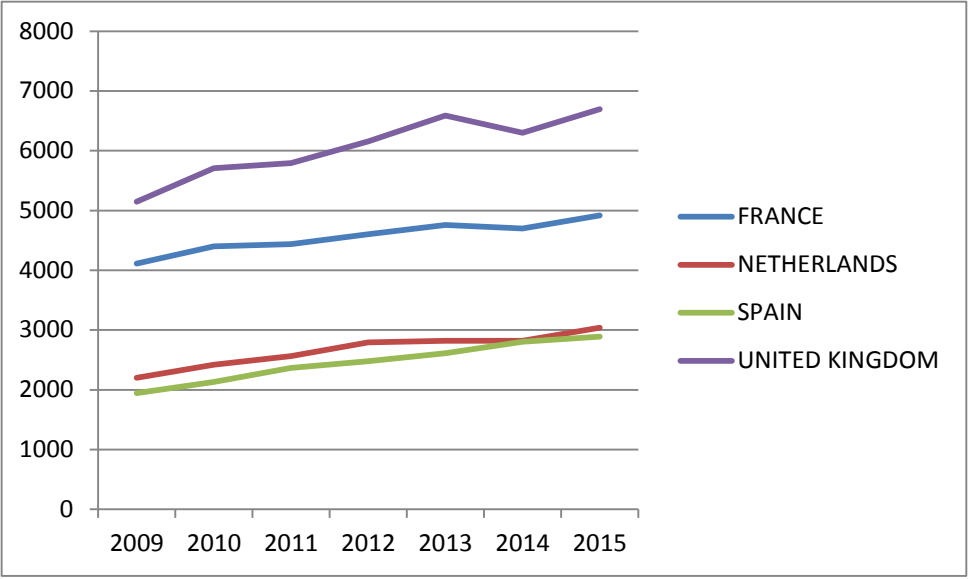

*Figure 1: Output numbers in Rare Diseases, four countries, 2009-2015*

In Figure 2, the shares in the output in rare diseases over a typology of five different funding sources is shown. The patterns of funding sources do vary among the four countries, as the overall composition for Great Britain shows a large share forthcoming from national funding in the rare diseases topic area. The Netherlands have the largest share of Other funding in rare diseases publications. Spain shows the smallest share in Other funding, on the hand shows the largest share in Europe-national funding in rare diseases. Finally, France shows the largest share in no funding in the publications on rare diseases.

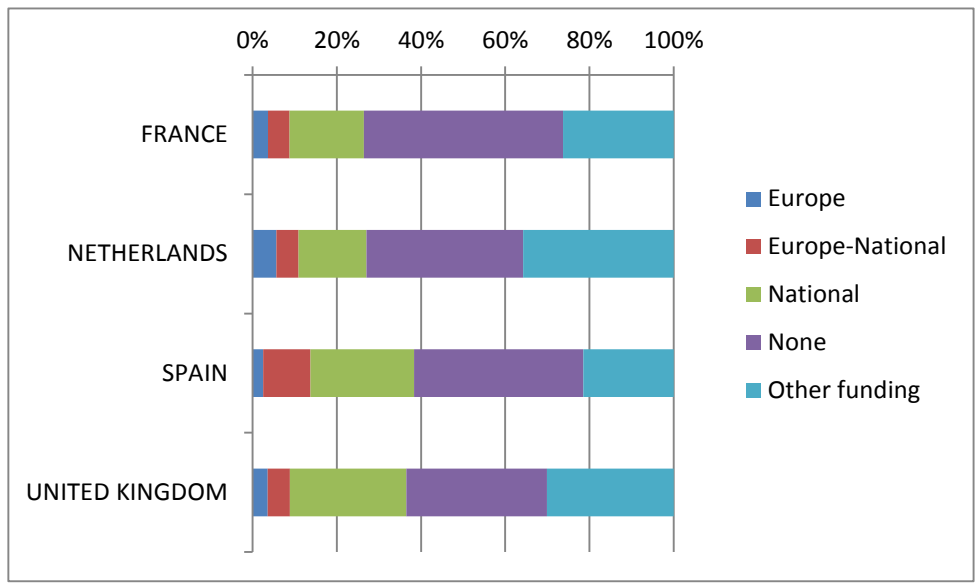

*Figure 2: Output shares in Rare Diseases by funding source, four countries, 2009-2015*

In Figure 3, the impact scores related to the rare diseases field are displayed. A first observation is that impact scores for all four countries show increasing patterns. A next observation is that we distinguish two patterns between the four countries, as Great Britain and the Netherlands show overall high impacts in research on rare diseases (with a more rapid increase for the former country), with relative lower impact levels for Spain and France (although approaching the worldwide average impact level).

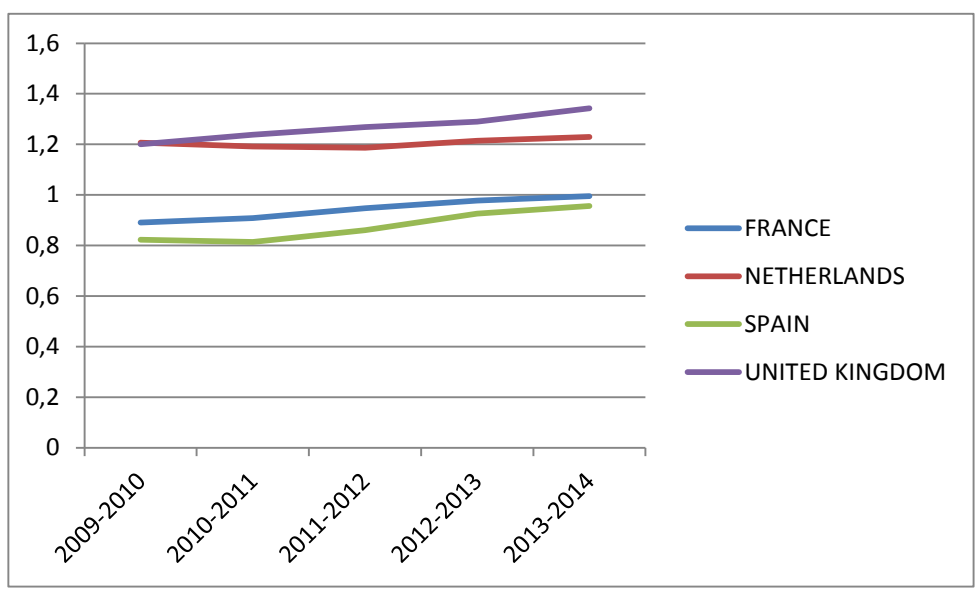

*Figure 3: Impact scores in Rare Diseases, four countries, 2009-2014/2015*

In Figures 4-7 we show parts a and b. In part a the output numbers per funding source are shown, while the b parts display the impact scores related to the output numbers shown in part a.

In Figures 4a and 4b the situation for France is depicted. In terms of output development, the none funded is decreasing, while the other types of funding sources show slow increases in numbers. The 'gap' between no funding, and the funded share of the French output is relatively large. When we shift our attention to the impact scores, the no funded part has the lowest impact, and is stable on that low level. The other four types of funding sources show higher impact level (fluctuating between 20% above worldwide average impact level to 60% above worldwide average impact level, for the Europe-national funding source).

Figures 5a and 5b display the output and impact for Great Britain. Contrary to France, we now observe some funding source types with higher numbers of publications (national and other funding). Europe and Europe-national show the lowest number of publications from Great Britain related to rare diseases research. Focusing on impact scores related to the output from Great Britain on rare diseases, we notice that the non-funded part has a low impact, similar to the situation found for France on this type, albeit it at a somewhat higher level. For Europe, Other funding, and nation, we observe impact level that fluctuate around 50% above average impact level for worldwide, while finally the output in Europe-national context displays a very high impact level, that is increasing to twice worldwide average impact level in 2013-2014/15.

In figures 6a and 6b, output and impact for the Netherlands are shown, in which we find some similarity to the situation described for Great Britain: decrease in output for no funding, increasing output for other funding and national (with some stabilization for the latter in the final stages of the period). Like for Great Britain, Europe-national and Europe have small output numbers. Impact wise, the no funding part has an average impact level, while we observe some strong fluctuation for the impact of in particular Europe-national and Europe (which is partially explained by the small numbers involved, which tends to occur more frequently in bibliometric analyses). Other funding and national shows more stable impact levels around 50% above worldwide average impact level.

Finally, in Figure 7a and 7b, the situation for Spain is displayed. In Figure 7a, the output is shown. Here we observe a somewhat different pattern as found for the other three countries, as here no funding is covering the largest part of the output, decreasing in numbers, while national as the second largest type of funding source is increasing in numbers. Other funding and Europe-national show increasing output numbers, while Europe has the lowest numbers of publications (and stable at that level). In terms of impact we find that no funding has again the lowest impact level, while the other four types of funding sources have higher impact scores, albeit somewhat more different as compared to the other three countries, still varying between on worldwide average impact to roughly some 60% above that worldwide average impact level.

*Figure 4a: Output numbers, France*  *Figure 4b: Impact scores, France*

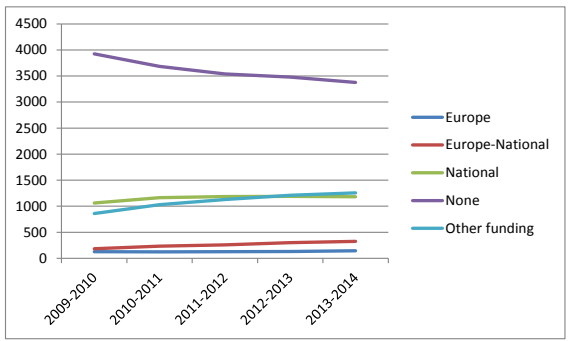 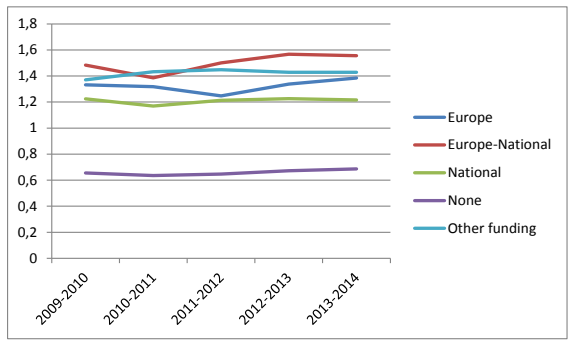

*Figure 5a: Output numbers, Netherlands*  *Figure 5b: Impact scores, Netherlands*

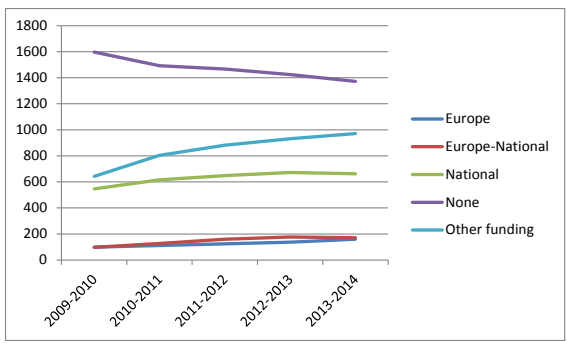 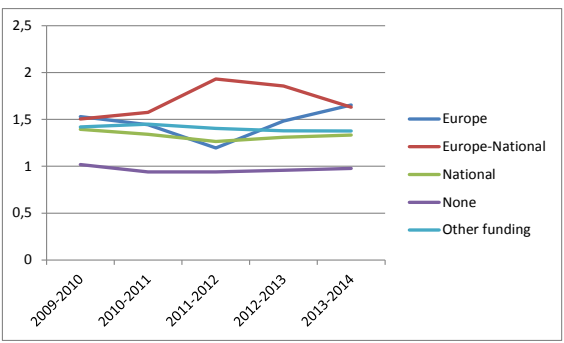

*Figure 6a: Output numbers, Spain*  *Figure 6b: Impact scores, Spain*

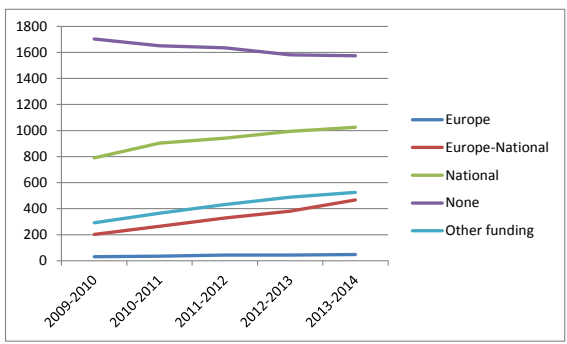 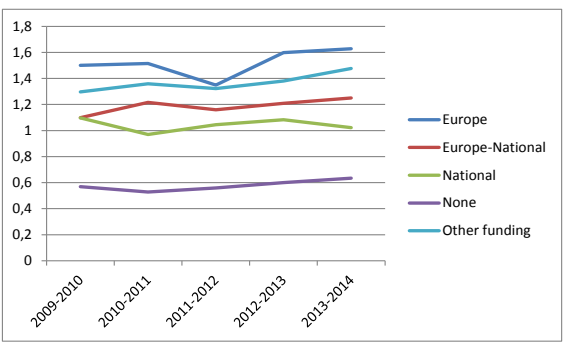

*Figure 7a: Output numbers, United Kingdom*  *Figure 7b: Impact scores, United Kingdom*

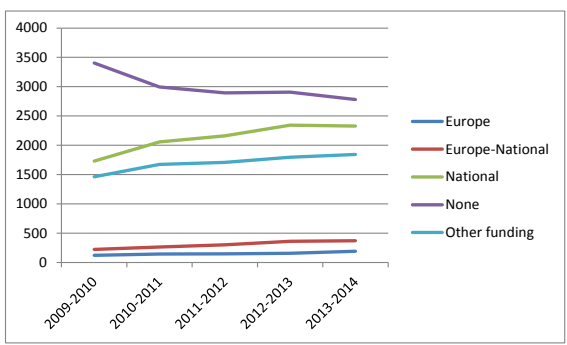 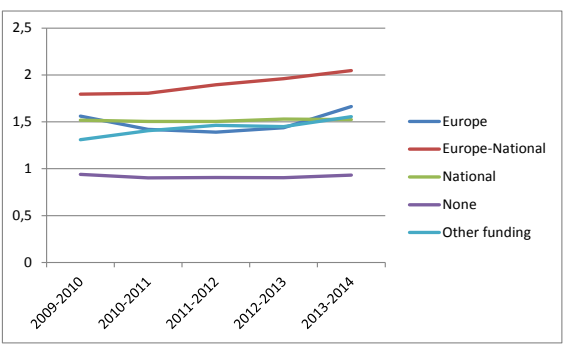

# Next steps to be taken in our analysis

Although to date theoretical claims about the influence of new agendas on research within scientific communities have not been measured using quantitative outcome data, bibliometrics provides a useful means to track interactions around topics of undone science within the science systems. In this paper we use funding acknowledgement information linked to scientific publications to consider the funding landscape of rare disease research across four EU Member States.

In this present version of the manuscript we have sorted rare disease publications from France, Netherlands, Spain and UK into four categories: no funding mentioned at all, European funding agency mentioned, national funding agency mentioned, and both EU and national funding mentioned. Currently these distinctions are not useful in and of themselves, as they are rather too heterogeneous to draw conclusions. However, they do at least provide an illustration of work-in-progress for which we hope to receive early-stage feedback.

In the subsequent rounds of cleaning and analysis we plan to make a more robust categorization of rare disease funders. This will include limiting our dataset to publications of these four member states associated with: a) the main national funding agencies, b) European Commission funding instruments (associated with Framework Programs), c) private companies and d) the significance of charities. On the basis of this classification we shall ask what share of the national publications are linked to the four categories, and which category is associated with the highest cited works on rare diseases across the countries.

The 'EU-funded' category will be constructed by searching for names of instruments associated with the Framework Programs 6 and 7 (EC, 2007). Following this sampling logic, we will exclude from the 'EU-funded' category names of instruments associated with the European Union, but not aimed (directly) towards supporting fundamental research (i.e. outside the Framework Programs), for example funding instruments linked to regional development goals or commercialization. Likewise we will exclude from the 'EU-funded category' funding organizations and schemes which are pan-European in their reach yet not associated with the European Commission, such as EMBO. This latter category has not been excluded from the 'European' category for the present analysis.

Although there is a clear accent placed on relevance by EU and national funding agencies, the logic of selecting high quality, leading scientific fundamental research via peer review persists. Indeed the push towards 'excellence' (in the case of some European and national instruments), and scientific quality more generally, is inscribed in funding agencies' attempts to build research capacity in rare diseases nationally and in Europe. In these latter instances citation output and impact-based analyses are therefore appropriate indicators. As such restricting European funding to Framework Program components will allow more meaningful interpretations of publication outputs and impacts than is presently afforded by the rather broad 'European funding' category.

Similarly by refining our 'national funding' category in later analysis to known funding agencies we hope to analyse publications which are supported by organisations whose mission is to promote continuation of fundamental research and breakthrough knowledge. We also hope to establish a category for national charities funding rare disease research in the four Member States. In producing this more precise classification, we hope to be able to interpret meanings of differences in citation scores with greater confidence than was allowed in the current provisional analysis. For instance, we can consider whether UK and Netherlands performing higher above world-average in terms of mean normalized citation scores is associated with different funding patterns in these respective states. This kind of analysis will probably also require further contextualizing information, such as theoretical accounts about the relative strength of public funding agencies and research charities in the different member states (c.f. Braun 1998, Whitley 2010).

Our hypothesis in the next stage of analysis is that European Commission funding initiatives are likely to be associated with the leading publications on rare diseases within the four member states (measured using normalized citation indicators). According to theoretical accounts, undone science topics articulated by social movements are most likely to gain traction in scientific communities through latching onto existing science policy instruments, like priority setting and funding (Hess, 2016). Indeed the provision of project and infrastructure support for rare diseases by funding agencies has been cited by activists and policymakers as a key means for building capacity on rare disease topics, both nationally and at European level (EUROPLAN, 2009). Targeting these conventional funding channels follows a logic that under-investment and lack of interest in rare diseases can be overcome by providing large amounts of research funding for attracting 'excellent' scientists to invest their attention and energy on hitherto marginal or 'unsexy' topics. Given the prestige and amounts of 'big science' funds attached to EU's instruments for funding fundamental research, we would therefore expect leading figures are likely to want to compete for these funds. Indeed links between large-scale project funds and leading science (in terms of impact on the field) often exhibit 'Matthew effect' dynamics (Langfeldt et al., 2015), especially in 'rapid discovery' fields like biomedical sciences costs for resources are high (Whitley, 2014). Finally this hypothesis is supported by an assumption that peer review by scientific elites is effective at identifying and selecting novel, vanguard research which will have wide impact on the scientific knowledge base. The evidence in support of this latter assumption is of course mixed – as many have used citation impact measures as a means to question the efficacy of peer review in anticipating novel, high impact work (Bornmann and Daniel, 2005). The provision of new empirical evidence to test this hypothesis is timely given the somewhat tentative nature of these claims.

Through these next steps we hope to arrive at a more fine-grained classification of the 'funding landscape' of rare disease research in EU member states, which allows us to compare the relative presence of different kinds of public and private funders in research outputs. This provides a potentially important means to validate empirically statements about the relative influence of different funders in the development of rare diseases, and illustrates the potential of bibliometrics as a tool for analyzing the dynamics of 'undone science' topics in science policy and the sociology of science.